\documentclass[]{spie}  

\usepackage{amsmath,amsfonts,amssymb}
\usepackage{graphicx}
\usepackage[colorlinks=true, allcolors=blue]{hyperref}
\usepackage{subcaption}
\usepackage{tablefootnote}
\usepackage{xcolor}
\definecolor{seagreen}{RGB}{46,139,87}

\title{Multi-sensor fusion for fine-guidance and milliarcsecond-level attitude estimation of balloon-borne telescopes}

\author[a,b]{Philippe Voyer}
\author[c,d]{Maya Amit}
\author[b]{Steven J. Benton}
\author[e,f,g]{Benjamin E. Boyd}
\author[h,i,j]{Anthony M. Brown}
\author[k]{Giulia Cerini}
\author[i]{Paul Clark}
\author[k]{Matthew Craigie}
\author[l]{Christopher J. Damaren}
\author[m]{Tim Eifler}
\author[n]{Spencer W. Everett}
\author[b]{Aurelien A. Fraisse}
\author[h,j]{Leo W.H. Fung}
\author[o]{Ajay S. Gill}
\author[b]{Suren Gourapura}
\author[d]{Eric Habjan}
\author[e]{John W. Hartley}
\author[p]{David Harvey}
\author[l]{Bradley Holder}
\author[k]{Eric M. Huff}
\author[j]{Mathilde Jauzac}
\author[b]{William C. Jones}
\author[j]{David Lagattuta}
\author[h,j]{Gavin Leroy}
\author[q]{Jason S.-Y. Leung}
\author[b]{Lun Li}
\author[b]{Thuy Vy T. Luu}
\author[h,i,j]{Richard Massey}
\author[c,d]{Jacqueline E. McCleary}
\author[e]{Adyn Miles}
\author[r]{Johanna M. Nagy}
\author[f,g]{C. Barth Netterfield}
\author[g]{Emaad Paracha}
\author[k,n]{Susan F. Redmond}
\author[k]{Jason D. Rhodes}
\author[k]{Andrew Robertson}
\author[e]{L. Javier Romualdez}
\author[d]{Sayan Saha}
\author[j]{Jürgen Schmoll}
\author[s,g]{Mohamed M. Shaaban}
\author[t]{Ellen Sirks}
\author[u]{Sut Ieng Tam}
\author[b]{Simon Tartakovsky}
\author[d,v,w,x]{Georgios N. Vassilakis}
\author[k]{André Z. Vitorelli}
\author[e,f,g]{Bryce Warren}
\author[y]{Alfredo Zenteno}

\affil[a]{Department of Mechanical \& Aerospace Engineering, Princeton University, 35 Olden St, Princeton, NJ, 08540, USA}
\affil[b]{Department of Physics, Princeton University, Jadwin Hall, Princeton, NJ, USA}
\affil[c]{Center for Data Science, New York University, 60 5th Ave, New York, NY 10011, USA}
\affil[d]{Department of Physics, Northeastern University, 360 Huntington Ave, Boston, MA, USA}
\affil[e]{StarSpec Technologies Inc., 402 Harmony Road, Units 8-11, Ayr, ON N0B 1E0, Canada}
\affil[f]{Dunlap Institute for Astronomy and Astrophysics, University of Toronto, 50 St. George Street, Toronto, ON M5S 3H4, Canada}
\affil[g]{Department of Physics, University of Toronto, 60 St. George Street, Toronto, ON, Canada M5R 2M8}
\affil[h]{Centre for Extragalactic Astronomy, Department of Physics, Durham University, Durham DH1 3LE, UK}
\affil[i]{Centre for Advanced Instrumentation (CfAI), Durham University, Durham DH1 3LE, UK}
\affil[j]{Institute for Computational Cosmology, Department of Physics, Durham University, South Road, Durham DH1 3LE, UK}
\affil[k]{Jet Propulsion Laboratory, California Institute of Technology, 800 Oak Grove Drive, Pasadena, CA, USA}
\affil[l]{University of Toronto Institute for Aerospace Studies (UTIAS), 4925 Dufferin Street, Toronto, ON, Canada}
\affil[m]{Department of Astronomy/Steward Observatory, University of Arizona, 933 N Cherry Ave, Tucson, AZ 85721, USA}
\affil[n]{Department of Physics, California Institute of Technology, 216 E California Blvd, Pasadena, CA 91125, USA}
\affil[o]{National Research Council Canada Herzberg Astronomy and Astrophysics Research Centre, 5071 West Saanich Rd, Victoria, BC V9E 2E7, Canada}
\affil[p]{Laboratoire d'Astrophysique, EPFL, Observatoire de Sauverny, 1290 Versoix, Switzerland}
\affil[q]{Department of Physics, Imperial College London, Prince Consort Rd, South Kensington, London SW7 2BW, UK}
\affil[r]{Department of Physics, Case Western Reserve University, Rockefeller Bldg., 10900 Euclid Ave, Cleveland, OH 44106, USA}
\affil[s]{Scale AI, Inc., 650 Townsend St, San Francisco, CA 94103, USA}
\affil[t]{Departamento de Física Teórica, Universidad Autónoma de Madrid, C. Nicolás Cabrera, 13-15, Fuencarral-El Pardo, 28049 Madrid, Spain}
\affil[u]{Institute of Physics, National Yang Ming Chiao Tung University, 1001 University Road, Hsinchu 30010, Taiwan}
\affil[v]{Institute of Astronomy, University of Cambridge, Madingley Rd, Cambridge CB3 0HA, UK}
\affil[w]{Department of Applied Mathematics and Theoretical Physics, University of Cambridge, Wilberforce Rd, Cambridge CB3 0WA, UK}
\affil[x]{Kavli Institute for Cosmology, University of Cambridge, Madingley Road, Cambridge CB3 0HA, UK}
\affil[y]{Cerro Tololo Inter-American Observatory / NSF NOIRLab, Vicuña, Coquimbo, Chile}

\authorinfo{Further author information: (Send correspondence to P.V.)\\P.V.: E-mail: pvoyer@princeton.edu}

\begin{document} 
\maketitle
\vspace{10pt}
\begin{abstract}
Balloon-borne telescopes rely on fine-guidance systems to achieve milliarcsecond image stability despite residual disturbances from the balloon environment. In these systems, the Fast Steering Mirror (FSM) stabilizes the image in two focal-plane axes, but leaves systematic, field-dependent residual motion induced by boresight roll. This effect, referred to as roll leakage, becomes more important for wider fields of view. In this work, roll leakage is characterized using data from the 2023 Superpressure Balloon-borne Imaging Telescope (SuperBIT) science flight. SuperBIT is a 0.5-m near-ultraviolet to near-infrared telescope that demonstrated milliarcsecond-level image stability during its 45-night science flight. We find that passive focal-plane star-camera measurements correlate strongly with independent roll measurements across a large set of science exposures, showing that boresight roll frequently drives residual focal-plane motion. We then develop a simulation framework combining optical ray tracing, asynchronous guide-star measurements, estimation, and FSM control to study this behavior. The framework is used to compare single-star and multi-star guidance architectures under realistic flight disturbances. For the SuperBIT geometry, we find that multi-star estimation reduces average roll-induced science-field image motion by $31.8\%$, increasing to $77.4\%$ for a representative geometry of GigaBIT, SuperBIT's planned larger-aperture successor. These results motivate further investigation of multi-star fine-guidance architectures for GigaBIT.

\end{abstract}

\keywords{Balloon-borne telescope, Fine Guidance System, Fast Steering Mirror, attitude estimation, image stabilization}

\clearpage
\setcounter{page}{1}
\pagestyle{plain}

\newpage
\section{INTRODUCTION}
\vspace{3pt}
\label{sec:intro}  

Balloon-borne telescopes promise space-quality imaging at a fraction of the cost of equivalent orbital missions, but their scientific return often depends on milli-arcsecond-level image stability, particularly for diffraction-limited optical observations \cite{Roth2025}. The balloon environment, however, introduces significant low-frequency pendulation and broadband structural vibrations that dominate the line-of-sight (LOS) disturbance budget. Operating in these conditions, the Gigapixel Balloon-borne Imaging Telescope (GigaBIT) is a 1.2-meter aperture, wide-field, high-resolution observatory designed for long-duration flights on NASA’s super-pressure balloon (SPB) platform \cite{Voyer2024}. GigaBIT inherits the dual-stage pointing architecture flight-proven by SuperBIT: a 0.5-meter near-infrared (NIR) to near-ultraviolet (NUV) telescope that flew four engineering flights and one science flight between 2015 and 2023 \cite{Romualdez2016, Romualdez2018, Romualdez2019, Gill2024}. Most recently, SuperBIT launched from Wānaka, New Zealand in April 2023 on a 45-night science flight, demonstrating $0.055 \pm 0.027$ arcsec ($1\sigma$) image stability over 300-second exposures \cite{Gill2024, Redmond2024, Saha2026}.

Both SuperBIT and GigaBIT combine a three-axis gimbal for telescope stabilization with an optical Fine Guidance System (FGS) that uses a Fast Steering Mirror (FSM) for jitter management and image stabilization. Each stage relies on a range of sensors for feedback, most notably three-axis gyroscopes and star cameras. For GigaBIT, the pointing requirement is sub-arcsecond three-axis telescope stabilization and $0.020$ arcsec ($1\sigma$) image stability over 300-second exposures. Meeting this requirement places strong demands on attitude estimation, sensor fusion, and control.

Recent advances in balloon-borne pointing technology have enabled high-precision three-axis attitude control from stratospheric platforms. In the last decade, instruments such as SuperBIT \cite{Gill2024}, FIREBALL-2 \cite{Hamden2020, CevallosAleman2024}, SUNRISE-III \cite{korpi-lagg2025, Berkefeld2026}, and PICTURE-C \cite{Mendillo2023}, which makes use of the Wallops Arcsecond Pointer (WASP) \cite{Stuchlik2015}, have demonstrated sub-arcsecond multi-axis control complemented by milli-arcsecond-level FSM fine guidance. SuperBIT’s gondola architecture has also been carried forward to EXCITE \cite{Nagler2022, Romualdez2024} and the proposed SCWI mission \cite{Miles2026}. Together, these missions have pushed balloon system pointing performance into a regime where residual focal-plane motion, rather than coarse telescope pointing, can become the limiting control problem.

A specific limitation of FSM-based fine guidance is that the FSM corrects image motion only in two focal-plane axes and cannot directly correct rotation orthogonal to the control axes. As a result, residual roll can be misinterpreted as tip-tilt image motion, stabilizing the guide star locally while leaving field-dependent residual motion elsewhere on the focal plane. Although SuperBIT proved that dual-stage LOS pointing can reach $\sim 0.050$ arcsec stability, its estimator was limited by (i) fine-pointing guide-star centroid latency and frame rate, (ii) reliance on a single guide star in the FGS, and (iii) structural flexure in the optical assembly that inertial sensors could not observe \cite{Voyer2024}. These limitations motivate a closer study of the resulting \emph{roll leakage}, how it couples into the FGS control loop, and strategies for reducing this ambiguity. Scaling to the GigaBIT instrument makes this problem more important. The longer focal length and lower structural resonant frequencies of the heavier optics increase sensitivity to sub-arcsecond pointing errors \cite{AMiles2026}. At the same time, GigaBIT’s fine-guidance cameras will benefit from the higher signal-to-noise ratio provided by the larger collecting area, allowing shorter exposure times and lower centroid latency. Unlike SuperBIT, where fine guidance relies on one camera at a time, GigaBIT’s larger focal plane can carry several low-latency guide cameras, providing faster feedback from a more representative sampling of the focal surface \cite{AMiles2026}. 

The contributions of this paper are threefold. First, we characterize roll leakage using data from the SuperBIT 2023 science flight and quantify its correlation with measured focal-plane motion and tip-tilt commands. Second, we develop a model of the FGS sensing and control problem that captures the coupling between boresight roll, guide-star location, centroid measurements, and FSM correction. Third, we use this model to study multi-guide-star estimation strategies for mitigating roll leakage in future balloon-borne telescopes, with direct application to the GigaBIT FGS design.\\

\section{Problem Formulation}
\label{sec:problem_formulation}
\vspace{3pt}

We consider the class of balloon-borne imaging telescopes that use a dual-stage pointing architecture, in which a coarse three-axis pointing system stabilizes the telescope body and a Fine Guidance System corrects residual image motion on the focal plane. The fine stage is assumed to use an FSM located in the optical path downstream of the telescope. The FSM provides two angular control inputs (tip and tilt), corresponding to focal-plane image motion in two axes, but it does not provide direct control authority about its normal axis.

The problem considered in this work is the coupling between residual boresight roll and focal-plane image motion in such a two-axis fine-guidance architecture. In particular, we are interested in the case where the fine-guidance loop locks onto a guide star measured by one focal-plane star camera (FSC), while other parts of the focal plane, including a second FSC and the science camera, remain sensitive to residual roll-induced motion.\\

\subsection{Pointing Architecture}
\label{subsec:pointing_architecture}
\vspace{2pt}

\begin{figure}[t] \centering \includegraphics[width=\linewidth]{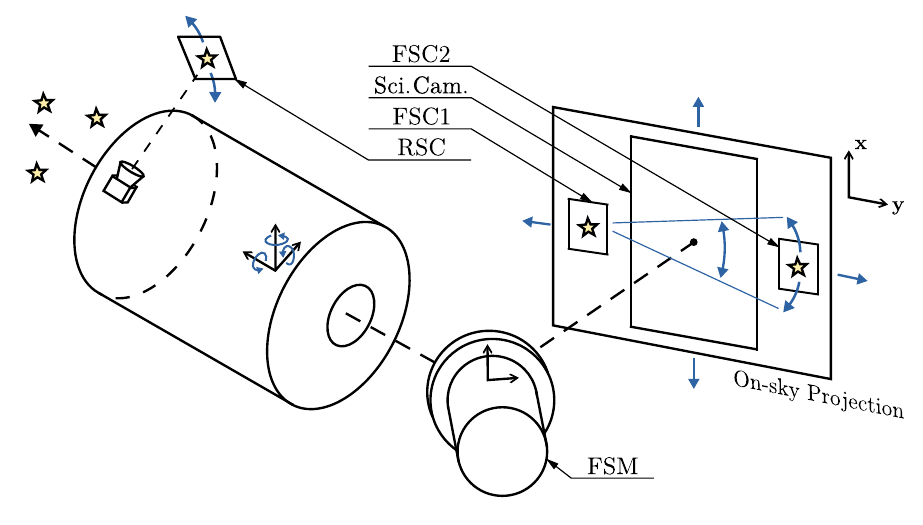} \caption{Representative (not-to-scale) diagram of the BIT fine-guidance architecture with the FGS active and locked on FSC1. The diagram shows the focal-plane star cameras (FSC1/FSC2), science camera, roll star camera (RSC), FSM, and the projection of telescope line-of-sight jitter (blue) onto the sky. FSC1 guide-star centroid measurements are used for fine pointing, while uncorrected roll-sourced motions can appear as residual focal-plane motion.} \label{fig:Fig1} \end{figure}

Figure~\ref{fig:Fig1} shows a representative diagram of the pointing architecture considered in this work. The telescope line of sight is first stabilized by the coarse pointing system, which controls the gondola and telescope attitude in three axes. Residual image motion is then corrected by the Fine Guidance System (FGS), which uses focal-plane guide-star measurements to command the FSM. The coarse pointing system operates independently of the FGS and relies on dedicated boresight and roll star cameras as well as gyroscopes for 'lost-in-space' attitude determination with sub-arcsecond telescope stabilization \cite{Romualdez2018}.

For SuperBIT, the FGS uses two focal-plane star cameras, denoted FSC1 and FSC2. During fine-guidance operation, the FGS selects a suitable guide star and uses the corresponding FSC as the active guidance sensor. The measured guide-star centroid is then used to define the fine-guidance error. Between camera updates, a Kalman filter propagates the line-of-sight estimate using axis-aligned gyroscope measurements. While this approach improves high-frequency estimation, post-flight analysis found that gyroscopes were a poor predictor of low-frequency focal-plane image motion, since no single gyro location fully captures the structural and optical dynamics that contribute to image motion at the focal plane \cite{Voyer2024}. The resulting line-of-sight estimate is converted into FSM tip and tilt commands, which steer the optical beam to reduce the measured guide-star motion \cite{Romualdez2018}. Camera parameters from the 2023 science flight are shown in Table \ref{tab:superbit_camera_specs}.

\begin{table}[!h]
\caption{SuperBIT camera specifications \cite{Gill2024}.}
\vspace{1pt}
\label{tab:superbit_camera_specs}
\centering
\renewcommand{\arraystretch}{1.05}
\begin{tabular}{lccc}
\hline
\rule{0pt}{2.2ex}
\textbf{Parameter} &
\textbf{FSC1} &
\textbf{FSC2} &
\textbf{RSC} \\
\hline
Sensor & Sony IMX392 & Sony ICX674 & Sony IMX264 CMOS \\
Sensor size [px] & $1920\times1200$ & $1940\times1460$ & $2448\times2048$ \\
Sensor size [mm] & $6.6\times4.2$ & $8.81\times6.63$ & $8.4\times7.1$ \\
Pixel size [$\mu$m] & $3.45\times3.45$ & $4.54\times4.54$ & $3.45\times3.45$ \\
Plate scale [arcsec px$^{-1}$] & 0.13 & 0.17 & 2.37 \\
Field of view [deg$^2$] & 0.003 & 0.006 & 2.2 \\
\hline
\end{tabular}
\end{table}

When one FSC is active, the other FSC can be treated as a passive focal-plane motion sensor when a suitable star is also present in its field of view. The active FSC measures the guide-star motion used by the FSM feedback loop, while the passive FSC measures residual motion at another location on the focal plane after the FSM correction has been applied. A practical benefit of this configuration is that the passive FSC can give a direct measure of the residual motion away from the stabilized guide star. If the FSM correction perfectly removed common translational jitter, both FSCs would show similarly reduced residual motion. If, instead, the FSM correction is partly compensating roll-induced motion about an off-axis guide star, the passive FSC can retain a correlated residual signal. This behavior is illustrated in Figure \ref{fig:Fig1}. This analysis, however, relies on simultaneous star acquisition on both FSCs, which is not guaranteed.\\

\subsection{Roll Leakage}
\label{subsec:roll_leakage}
\vspace{2pt}

We define \emph{roll leakage} as the residual field rotation induced by boresight roll after the FSM has stabilized the active guide star.  The quantity of interest is not only the angular roll motion of the telescope, but the way that this roll motion maps to the focal plane as field rotation after two-axis fine-guidance correction. We also distinguish between gimbal roll and boresight roll: the former is defined about the intermediate gimbal axis, while the latter is defined about the telescope boresight after the Euler rotation sequence. Roll leakage should not be confused with field-dependent projection effects introduced by the FSM, in which tip-tilt corrections produce nonlinear image motions across the focal plane as a function of FSM orientation \cite{Hilkert2014}.

For a small boresight roll perturbation, a point on the focal plane is displaced in proportion to its distance from the field rotation center. Roll does not produce a uniform translation, but instead induces position-dependent image motion across the field. A two-axis FSM correction can remove the apparent displacement at one selected location, namely the active guide star, but it cannot remove the rotational component of the motion over the full focal plane. This has an important consequence for the FGS. If residual boresight roll moves the active guide star, the fine-guidance loop interprets that motion as a two-axis centroid error. The FSM, having no direct way to separate the two, corrects both. The guide star is therefore stabilized locally. However, this correction effectively pins the focal-plane motion at the guide-star location. Away from that location, the remaining displacement depends on the separation from the guide star and from the effective field rotation center. In this sense, the FSM correction changes the apparent center of residual roll motion from the optical axis to the active guide star. The result is a local/global tradeoff. The active FSC can show small residual centroid motion because it is directly controlled by the FSM (good locally), while a passive FSC or the science camera can still experience residual image motion (bad globally). During a long science exposure, this residual motion will broaden or elongate the point-spread function (PSF), even when the active guide star appears well stabilized. This effect naturally becomes more pronounced for larger focal planes, such as that of GigaBIT \cite{AMiles2026}.

Finally, roll leakage depends directly on the performance of the coarse pointing system. With perfect gimbal roll control, this leakage mechanism would be mostly eliminated. Improving coarse roll is therefore complementary to the fine-guidance approach considered here, and remains a primary path for reducing this type of motion.\\

\subsubsection{SuperBIT Experimental Evidence}
\vspace{2pt}

\label{subsec:superbit_experimental_evidence}
\begin{figure}[t]
    \centering
    \includegraphics[width=0.498\linewidth]{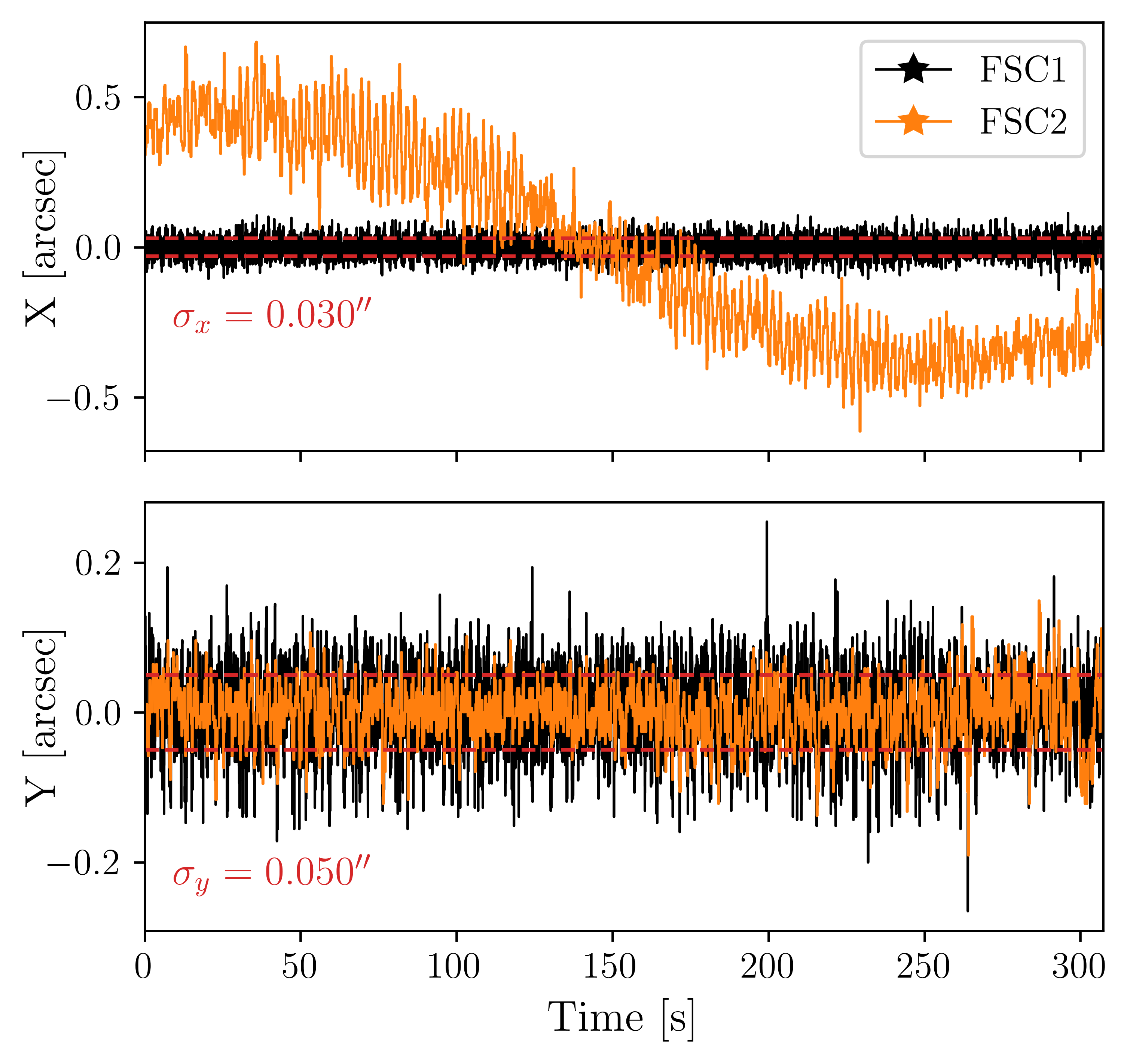}\hspace{0.004\linewidth}%
    \includegraphics[width=0.498\linewidth]{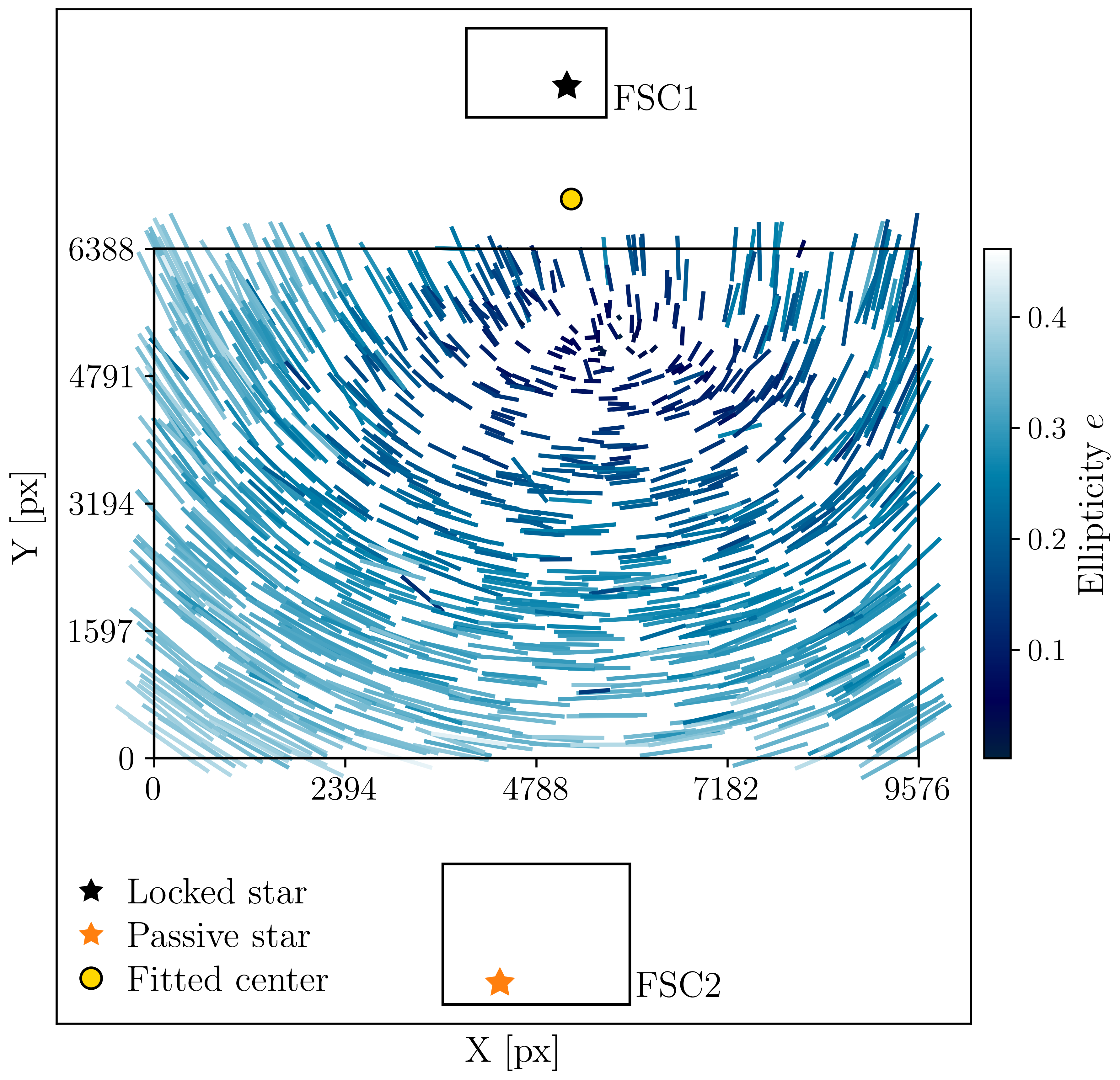}
    \caption{Representative $\sim300$ s SuperBIT exposure of Abell 3411 taken on 03-05-2023, at payload latitude $-46.0103^\circ$ and longitude $142.1318^\circ$, displaying roll jitter leakage on the focal plane. Left: pitch and cross-pitch focal-plane star-camera centroid motion for FSC1 and FSC2, with the FGS locked on FSC1. Right: Science-camera PSF ellipticity field overlaid on a to-scale focal-plane layout, with ellipse major axes scaled by a factor of 7000 for visibility. The ellipses show the locations of FSC1, FSC2, the locked and passive guide stars, and the field rotation center fitted from the PSF ellipticities.}
    \label{fig:figure2}
\end{figure}

The SuperBIT 2023 science flight provides a useful dataset for identifying this effect. SuperBIT observed 40 galaxy-cluster targets, including both complete and partial observations, across roughly 5081 stable five-minute science exposures. During fine-guided science observations, one FSC was used in the FGS feedback loop, while the second FSC could act as a passive sensor when a sufficiently bright star was present in its field of view. In these cases, the passive FSC recorded centroid motion without directly driving the FSM, providing an independent measurement of residual focal-plane motion after stabilization at the active guide star.

Figure~\ref{fig:figure2} shows a representative $\sim300$~s exposure of Abell~3411 where roll disturbance was significant. In this exposure, the FGS is locked on FSC1, which remains locally stabilized by the FSM feedback loop, while FSC2 acts as the passive focal-plane sensor. The left panel shows the measured $x$ and $y$ centroid motion on both FSCs. Although the active guide star is stabilized locally, the passive FSC records residual motion at another location on the focal plane. The right panel shows the corresponding science-camera PSF ellipticity field scaled and overlaid on the focal-plane layout, together with the guide-star locations and the fitted field rotation center of the science camera. The spatial structure of the ellipticity field is consistent with residual roll-induced motion across the focal plane. The fitted field rotation center also suggests that some of the $y$-direction motion is not fully corrected in this specific exposure.

\begin{figure}[t]
    \centering
    \includegraphics[width=0.65\linewidth]{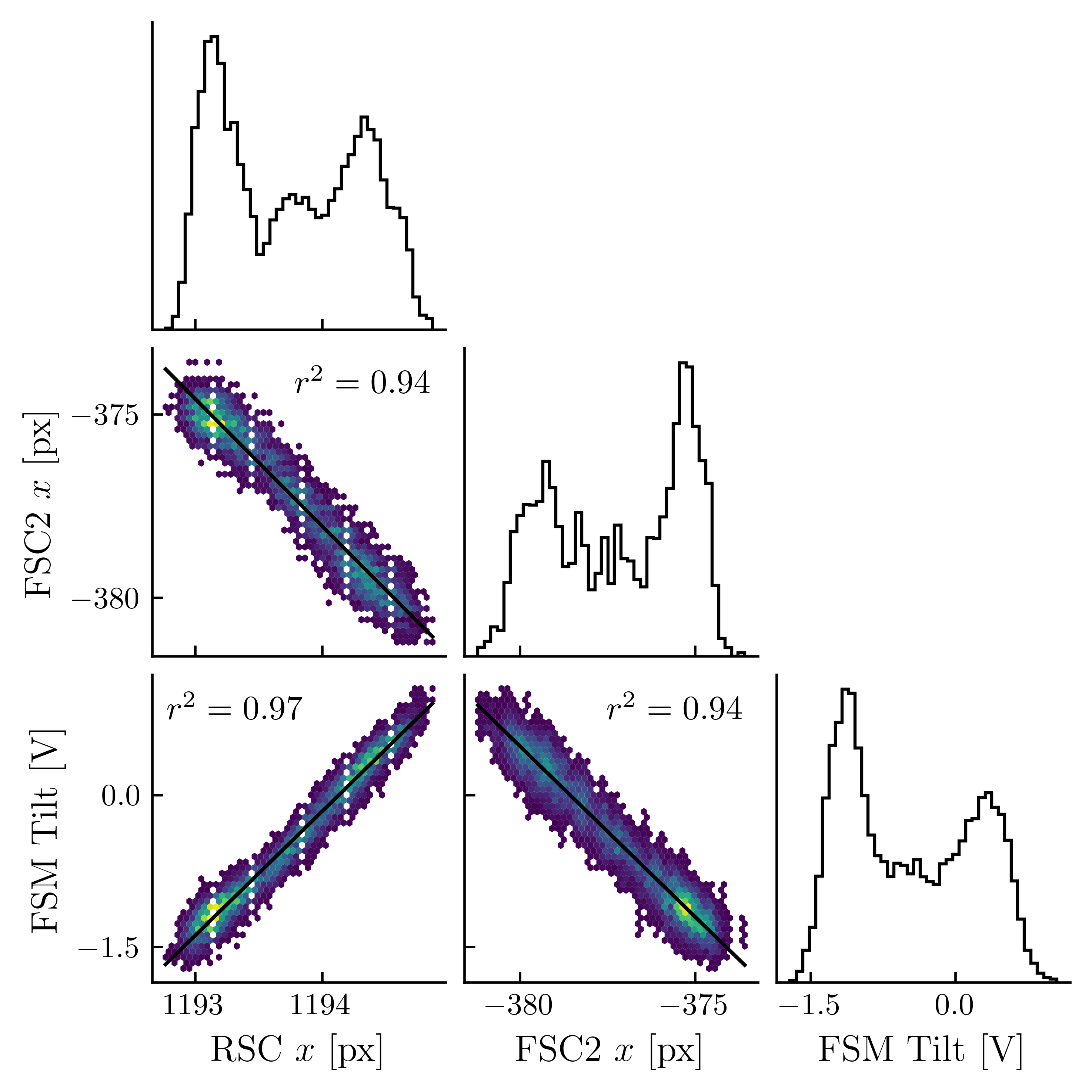}
    \caption{Triangle plot of a $\sim300$ s SuperBIT exposure of Abell 3411 taken on 03-05-2023. Pairwise correlations between FSC2 $x$ centroid motion, RSC $x$ centroid motion, and FSM tilt command are shown. Vertical gaps in the camera bins are an artefact of the centroiding algorithm discretizing the reported centroid positions. The best fit line for each box is shown in black.}
    \label{fig:corner_plot}
\end{figure}

We also use SuperBIT data to test whether residual motion on the passive FSC is consistent with roll leakage in most exposures. In particular, passive FSC centroid motion is compared with independent measurements of roll motion from the roll star camera (RSC), as well as FSM commands. A correlated component between passive FSC motion and the measured roll signal indicates that a large part of the residual focal-plane motion is consistent with boresight roll rather than purely translational jitter. Figure \ref{fig:corner_plot} shows that, for the same Abell 3411 exposure, the bimodal roll signal measured by the RSC is reproduced in the passive FSC2 centroid motion. This indicates that roll-induced disturbances are directly mapped onto the focal plane as residual image motion. Because the FSM acts to suppress this motion at the active guide star, the commanded FSM signal exhibits the same bimodal structure.

\begin{figure}[t!]
    \centering
    \includegraphics[width=\linewidth]{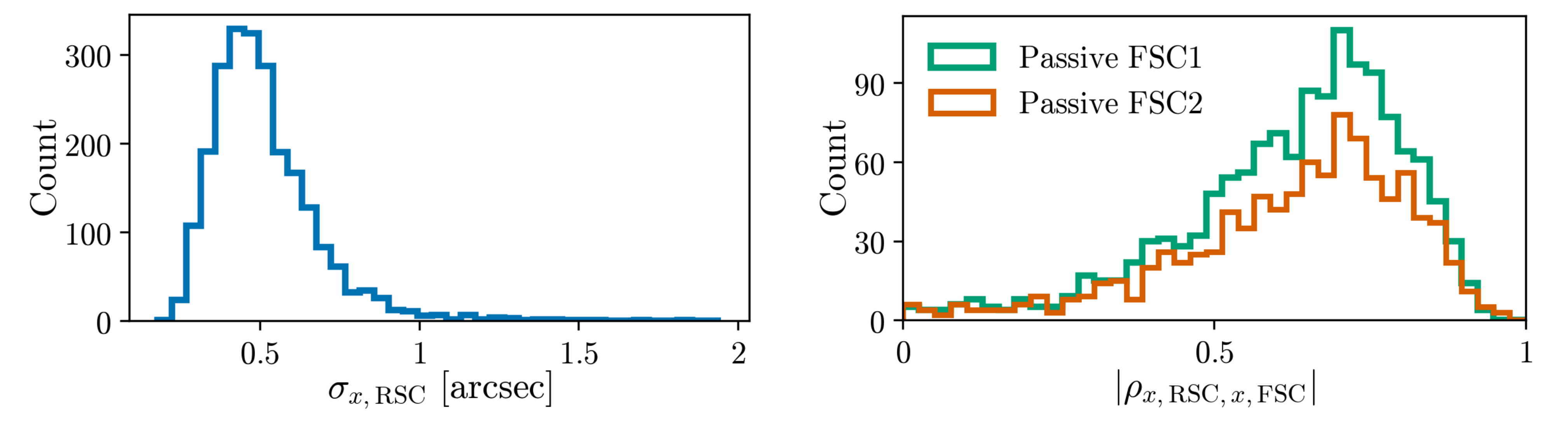}
    \caption{Left: Distribution of RSC standard deviation for good two-guide-star exposures ($n=2348$). Right: Distribution of the absolute correlation between passive FSC centroid motion and RSC centroid motion from Eq. \ref{eq:corr} for the same exposures, separated by whether FSC1 ($n=1379$) or FSC2 ($n=969$) is passive.}
    \label{fig:histograms}
\end{figure}

While Figures~\ref{fig:figure2} and \ref{fig:corner_plot} illustrate a single representative exposure, the same analysis was repeated across all science exposures with the FGS operating in closed loop and with valid guide-star detections on both FSCs. We retain only exposures satisfying
\begin{equation}
\max_i \Delta x_{\mathrm{FSC}i} < 10~\mathrm{px},
\label{eq:flag1}
\end{equation}
where $\Delta x_{\mathrm{FSC}i}$ denotes the centroid range during the exposure, excluding exposures with very large guide-star motions or loss of lock. We further restrict the sample to exposures meeting the image-quality criteria of the current SuperBIT science pipeline \cite{Saha2026}. The resulting dataset comprises 2348 of the 5081 exposures acquired during the SuperBIT flight.

The results across all selected exposures are summarized in Figure \ref{fig:histograms}. The left panel indicates that the typical roll-sourced image motion is of order $0.5''$, indicative of the good performance of the coarse roll stage. The right panel shows the distribution of absolute correlation coefficients between passive FSC centroid $x$ motion and the corresponding RSC roll measurement. The Pearson correlation coefficient is computed as
\begin{equation}
\rho_{x,\mathrm{RSC},x,\mathrm{FSC}}
=
\frac{
\mathrm{cov}\!\left(x_{\mathrm{RSC}},x_{\mathrm{FSC}}\right)
}{
\sigma_{x,\mathrm{RSC}}\ \sigma_{x,\mathrm{FSC}}
}.
\label{eq:corr}
\end{equation}
The distribution is skewed toward high correlation coefficients, with most exposures having $|\rho_{x,\mathrm{RSC},x,\mathrm{FSC}}| > 0.5$. This indicates that image motion measured by the passive FSC is frequently correlated with boresight roll measured by the RSC, providing evidence that roll disturbances commonly leak onto the focal plane during fine guidance. Since most exposures exhibit sub-arcsecond roll motion, this behavior is not limited to a small number of high-roll events, but is observed throughout the flight data, suggesting that roll leakage is an inherent feature of the fine-guidance architecture.\\

\subsubsection{Mitigation}
\vspace{2pt}

Mitigating roll leakage is therefore a key consideration for wide-field telescopes with fine guidance systems. The effect is observed across a large fraction of SuperBIT science exposures and persists even when the residual roll motion is relatively small. This motivates the investigation of guidance architectures aimed at mitigating roll leakage in wide-field telescopes that rely on local guide-star stabilization.

A natural way to mitigate roll leakage is to estimate focal-plane motion using measurements from multiple guide stars distributed across the field of view rather than relying on a single local guide star. Because roll produces position-dependent image motion across the focal plane, simultaneous measurements from multiple locations provide information that can be used to separate rotational and translational disturbances. Similar approaches have been used in space-borne observatories; for example, the Hubble Space Telescope historically relied on two guide stars to constrain roll during fine pointing \cite{Rafelski2026}. Although gyroscopes can improve high-frequency estimation between camera updates, SuperBIT flight data suggest that inertial measurements alone do not fully capture the structural and optical disturbances responsible for focal-plane image motion \cite{Voyer2024}. Additional focal-plane measurements are therefore likely required to mitigate roll leakage.

A primary challenge for multi-star estimation, however, is guide-star availability. Multiple stars must be present within the guide sensors' field of view and be sufficiently bright to provide reliable centroid measurements at the required update rate. This constraint is particularly restrictive for narrow-field instruments, and was one of the practical limitations of the SuperBIT architecture. GigaBIT alleviates this challenge through its substantially larger focal plane and collecting area, providing access to more and fainter guide stars during a typical observation \cite{AMiles2026}. This enables the investigation of estimation architectures that make use of multiple simultaneous focal-plane measurements. The simulation framework developed in this work is intended to evaluate such architectures and quantify their impact on focal-plane image stability.\\

\section{Simulation Framework}
\vspace{3pt}

\begin{figure}[t!]
    \centering
    \includegraphics[width=0.96\linewidth]{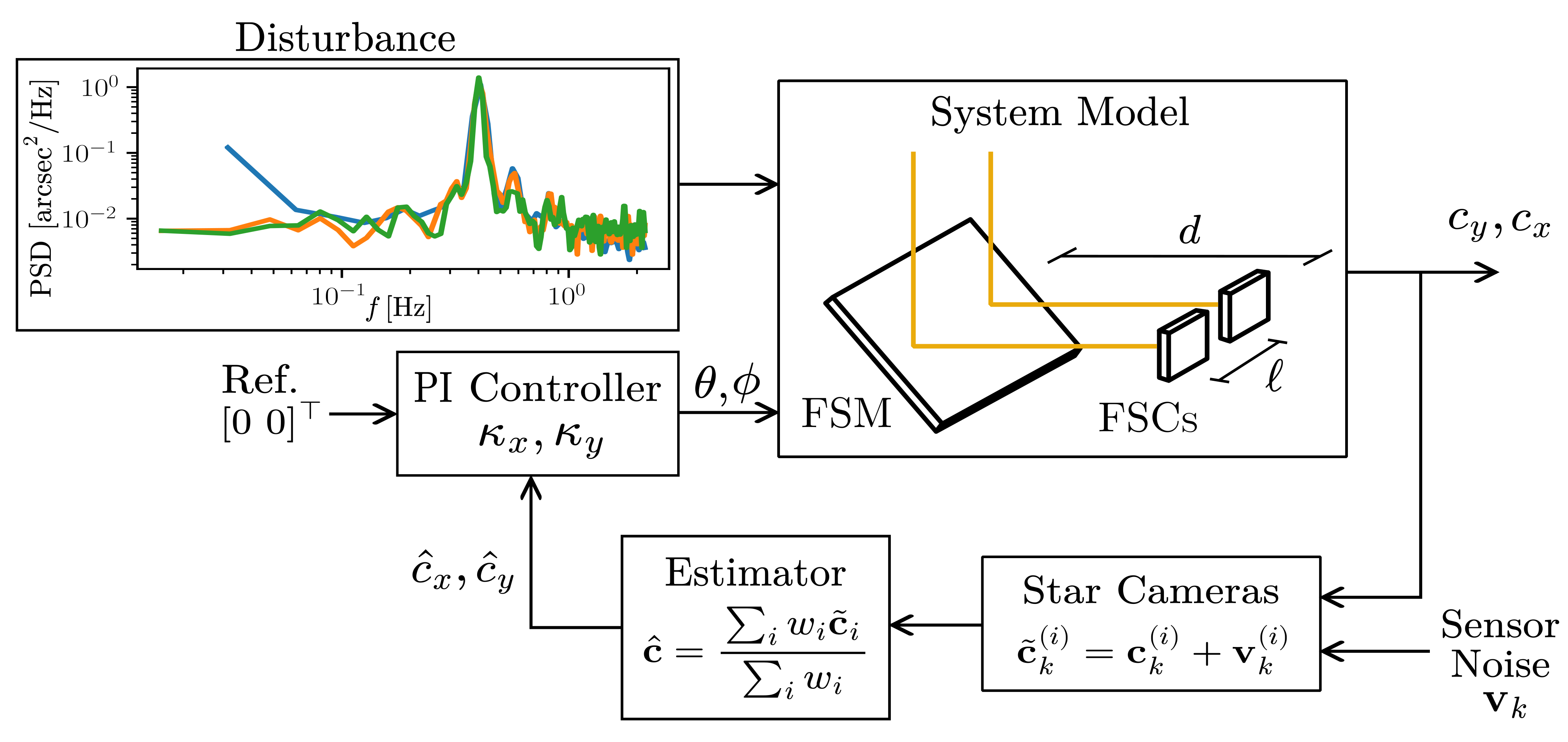}
    \caption{Block diagram of the simulated Fine Guidance System with multi-star estimation and FSM feedback. The disturbance plot is from representative roll and boresight star camera (RSC, BSC) power spectral densities (psd) from the 2023 SuperBIT flight. The FGS model is parametrized by distances $d$ and $\ell$.}
    \label{fig:sim_block_diagram}
\end{figure}

We simulate the Fine Guidance System as a closed-loop feedback system consisting of disturbance, process, measurement, estimation, and control components, as shown in Figure \ref{fig:sim_block_diagram}. The framework propagates telescope jitter through an optical ray-tracing model, generates asynchronous focal-plane measurements, estimates line-of-sight motion, and computes FSM commands for image stabilization. The following subsections describe the implementation of each component and the assumptions used in the simulation.\\

\subsection{Process Model}

In science observation mode, the gondola is stabilized to sub-arcsecond levels, with residual jitter in pitch, cross-pitch, and boresight roll reaching the FGS. We define a nominal optical configuration in which the guide-star rays intersect the centers of their respective FSC focal planes. For each FSC, the corresponding nominal input ray is obtained by backward ray tracing from the focal-plane center through the optical system to the telescope entrance pupil. These nominal rays define the reference line of sight for each guide star and are then perturbed by telescope jitter. The simulation tracks the centroid motion of all asynchronous FSCs simultaneously, allowing individual cameras to be designated as active, passive, or estimator inputs depending on the guidance architecture under consideration.

We use the ray-tracing model from Voyer et al. \cite{Voyer2024, Voyer2024thesis} to propagate each FSC ray through the FSM and onto the focal plane. The simulation takes in a given optical focal plane configuration and the SuperBIT or GigaBIT FGS via the parameters $d$ and $\ell$ in Figure \ref{fig:sim_block_diagram}. It then runs through the closed-loop with FSM control and estimation mode. While end-to-end telescope dynamics simulations are also under development \cite{Boyd2026}, the present work focuses on the fine-guidance system, for which an optical model is sufficient.

\noindent As outlined in Voyer et al.\cite{Voyer2024, Voyer2024thesis}, the FSM is presented as a plane with body-frame normal vector 
\begin{equation}
    \mathbf{n}_b = \mathbf{C}_{b,m}(\alpha, \beta)\ \mathbf{n}_m(\theta, \phi),
\end{equation}
\noindent where $\alpha$ and $\beta$ are FSM mounting angles, and with 
\begin{equation}
    \mathbf{n}_m(\theta, \phi)=\begin{bmatrix}\cos\theta \cos\phi \\ \cos\theta \sin\phi \\-\sin\theta\cos\phi\end{bmatrix},
\end{equation}
with $\theta$ and $\phi$ as tip and tilt, respectively, assumed very small. For a ray with origin $\mathbf{p}_r$ and direction $\mathbf{r}_i$, the plane intersection is 
\begin{equation}
    \mathbf{p}_s = \mathbf{p}_r +\frac{\mathbf{n}_b^\top(\mathbf{p}_0-\mathbf{p}_r)}{\mathbf{n}_b^\top \mathbf{r}_i}\mathbf{r}_i,
\end{equation}
\noindent and the reflected direction is 
\begin{equation}
    \mathbf{r}_r=\begin{pmatrix}\mathbf{1}-2\mathbf{n}_b\mathbf{n}_b^\top\end{pmatrix} \mathbf{r}_i. 
\end{equation}
For a single surface configuration, the optical path is completed by intersecting $\mathbf{r}_r$ with the focal plane and projecting that point into the plane.\\

\subsection{Measurement \& Estimation Models}
\label{sec:Measurement_Estimation}
The Fine Guidance System estimates focal-plane image motion using centroid measurements from the focal-plane star cameras (FSCs). Each FSC measures the displacement of its guide star relative to the nominal center of its focal plane. Let $\mathbf{c}_i(t)\in\mathbb{R}^2$ denote the noiseless centroid displacement measured by camera $i$. When camera $i$ produces a measurement at time $t_k^{(i)}$, the reported centroid is

\begin{equation}
\tilde{\mathbf{c}}_i\!\left(t_k^{(i)}\right)
=
\mathbf{c}_i\!\left(t_k^{(i)}\right)
+
\mathbf{v}_i\!\left(t_k^{(i)}\right),
\end{equation}

where $\mathbf{v}_i\sim\mathcal{N}(0,\boldsymbol{\Sigma}_v)$ is zero-mean Gaussian centroid noise. The FSCs operate asynchronously, with update rates drawn from $f_i\sim\mathcal{N}(\bar{f}_i,\sigma_{f,i}^2)$, where $\bar{f}_i$ and $\sigma_{f,i}$ are user-defined parameters. Between updates, the most recent centroid measurement is held constant, so the estimator receives a zero-order-hold measurement stream. Specific simulation parameters are summarized in Section \ref{sec:Numerical_results}.\\

The simulation tracks centroid motion from all FSCs simultaneously, allowing different guidance architectures to be evaluated using the same disturbance realization and control law. Two estimation modes are considered:

\begin{itemize}
    \item \textbf{Single-star}: representative of the SuperBIT fine-guidance architecture, in which a single guide star is used for feedback throughout the exposure. The centroid estimate is taken directly from the active FSC,
    \begin{equation}
    \hat{\mathbf{c}}_k =
    \tilde{\mathbf{c}}_{\mathrm{active},k}.
    \end{equation}

    \item \textbf{Multi-star}: measurements from all available FSCs are fused to form a common estimate of focal-plane motion,
    \begin{equation}
    \hat{\mathbf{c}}_k
    =
    \frac{\sum_i w_i\,\tilde{\mathbf{c}}_{i,k}}
         {\sum_i w_i},
    \end{equation}
    where the weights are chosen as
    \begin{equation}
    w_i
    =
    \tilde{w}_i
    \exp\!\left(
    -\frac{\tau_i}{\tau_d}
    \right),
    \label{eq:weights}
    \end{equation}
    with $\tilde{w}_i$ a nominal camera weight, $\tau_i$ the age of the most recent measurement from camera $i$, and $\tau_d$ a user-defined decay constant.
\end{itemize}

More recent measurements therefore contribute more strongly to the fused estimate than older measurements. While measurement age is used here for simplicity, quantities more directly tied to measurement quality, such as guide-star brightness, may provide a better weighting metric for implementation. The nominal camera weights could likewise be selected based on the focal-plane geometry, for example to emphasize guide stars located closer to the science sensors. Overall, we assess estimator performance through the residual image motion over the science camera after FSM correction. For the present work, we model only two FSCs, as shown in the System Model block of Figure \ref{fig:sim_block_diagram}.

Finally, gyroscope measurements are not included in the estimators considered here. Usefully integrating gyro information and faster focal-plane sensing are left for future work.\\

\subsection{Control Model}

Next, we command the FSM using independent proportional-integral (PI) controllers in the focal-plane $x$ and $y$ directions. The nominal optical configuration is defined such that the guide-star rays intersect the centers of their respective FSC focal planes. The control objective is therefore to regulate the centroid coordinates, expressed in the local FSC frame, to the reference
\begin{equation}
\mathbf{c}_{\mathrm{ref}} =
\begin{bmatrix}
0 \\ 0
\end{bmatrix}.
\end{equation}
The commanded FSM angles are computed as
\begin{equation}
\theta_k
=
\kappa_{p,y}\hat{c}_{y,k}
+
\kappa_{i,y}
\sum_{j=0}^{k}
\hat{c}_{y,j}\Delta t,
\end{equation}
\begin{equation}
\phi_k
=
-\kappa_{p,x}\hat{c}_{x,k}
-
\kappa_{i,x}
\sum_{j=0}^{k}
\hat{c}_{x,j}\Delta t,
\end{equation}
where $\kappa_{p,x}$ and $\kappa_{p,y}$ are proportional gains, $\kappa_{i,x}$ and $\kappa_{i,y}$ are integral gains, and $\Delta t$ is the simulation time step. The sign convention follows the local sensitivity of the FGS geometry.

The FSM response is modeled as a first-order system,
\begin{equation}
\tau\dot{\mathbf{u}}(t)
+
\mathbf{u}(t)
=
\mathbf{u}_{c}(t),
\end{equation}
where $\mathbf{u}_c$ is the commanded FSM angle vector, $\mathbf{u}$ is the applied FSM angle vector, and $\tau$ is the actuator time constant.\\

\subsection{Disturbance Model}
\begin{figure}[!h]
    \centering
    \includegraphics[width=0.6\linewidth]{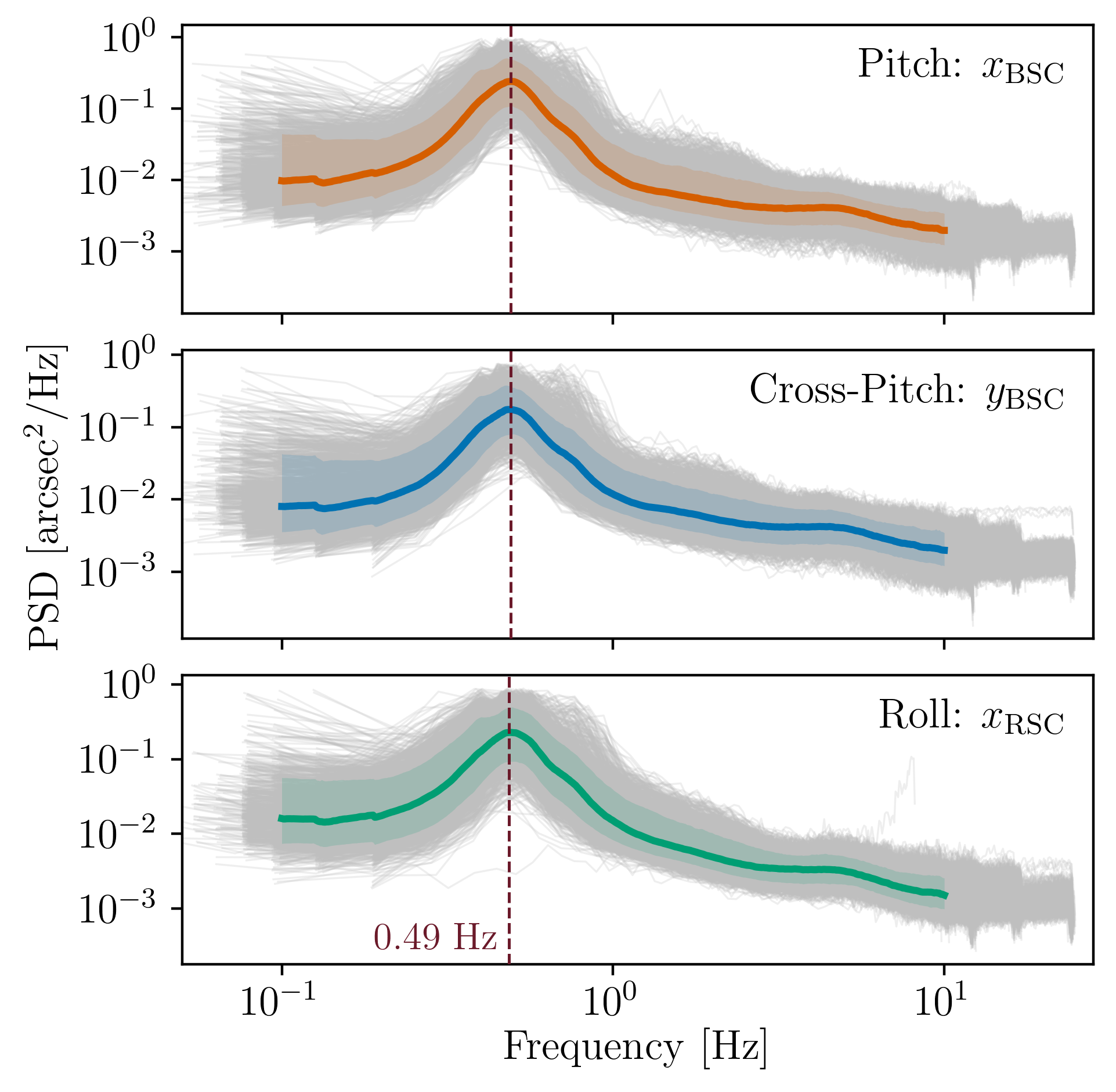}
    \caption{Power spectral densities of boresight and roll star camera (BSC, RSC) image motion measured during SuperBIT science observations. Gray curves show individual exposures, while colored curves show the median PSD, with halo regions indicating the 10th--90th percentile range.}
    \label{fig:superbit_psd}
\end{figure}

To inject representative disturbances, we add 3-axis line-of-sight jitter to the nominal input rays. To estimate the disturbance environment entering the FGS, we use flight data from the SuperBIT 2023 science mission and compute the per-axis power spectral density (PSD) of the measured image motion. While the exact disturbance spectrum depends on the mechanical and optical configuration of a given balloon payload, the SuperBIT data provide a representative example of the vibrational environment encountered by stratospheric balloon observatories. AAs noted in previous studies, balloon-borne telescopes exhibit prominent low-frequency pendulation modes arising from the balloon–flight train system \cite{Roth2025, Romualdez2018, Kassarian2021, Kassarian2024}. Higher-frequency disturbances, including contributions from onboard mechanisms, structural vibrations, and reaction wheels, are also present and are primarily captured by high-rate gyroscopes and accelerometers, but are outside the scope of this section. Figure \ref{fig:superbit_psd} shows all three axes exhibiting a dominant peak near $0.49\ \mathrm{Hz}$, corresponding to the primary pendulation mode of the observatory. While this feature dominates the median PSD, individual exposures often exhibit multiple peaks in the sub-$1\ \mathrm{Hz}$ regime associated with higher-order pendulation modes and changing flight conditions. An example single-exposure PSD is shown in the disturbance block of Figure \ref{fig:sim_block_diagram}, where a dominant low-frequency peak is accompanied by several smaller peaks. The broad feature in Figure \ref{fig:superbit_psd} therefore likely reflects a combination of several closely spaced low-frequency modes rather than a single oscillation.

To incorporate this into the simulation, the disturbance is modeled as a small rigid-body attitude perturbation applied to the nominal input-ray bundle. Let $\mathbf{C}_d(t)\in\mathrm{SO}(3)$ denote the disturbance rotation, where $\mathrm{SO}(3)=\{\mathbf{C}\in\mathbb{R}^{3\times3}\mid \mathbf{C}^{\top}\mathbf{C}=\mathbf{I},\ \det(\mathbf{C})=1\}$ is the group of three-dimensional rotation matrices. The disturbance is parameterized by the small-angle vector
\begin{equation}
    \delta\boldsymbol{\theta}_d(t)
    =
    \begin{bmatrix}
        \delta\theta_{x,d}(t) &
        \delta\theta_{y,d}(t) &
        \delta\theta_{z,d}(t)
    \end{bmatrix}^{\top},
\end{equation}
representing pitch, cross-pitch, and boresight-roll disturbances, respectively. The disturbance is applied to the nominal input-ray bundle through the common rotation
\begin{equation}
    \mathbf{r}_i(t)=\mathbf{C}_d(t)\mathbf{r}_{i,0},
\end{equation}
and
\begin{equation}
    \mathbf{p}_i(t)=\mathbf{p}_b+\mathbf{C}_d(t)\bigl(\mathbf{p}_{i,0}-\mathbf{p}_b\bigr),
\end{equation}
for all cameras $i$, where $\mathbf{r}_{i,0}$ and $\mathbf{p}_{i,0}$ are the nominal ray direction and ray origin, respectively, and $\mathbf{p}_b$ is the boresight reference point. This applies a common rigid-body rotation to the entire input-ray bundle about the boresight axis. The disturbance rotation is generated from the small-angle disturbance vector $\delta\boldsymbol{\theta}_d(t)$ through the exponential map,
\begin{equation}
    \mathbf{C}_d(t)
    =
    \exp\!\left(
    \delta\boldsymbol{\theta}_d(t)^\times
    \right),
\end{equation}
where $(\cdot)^\times$ denotes the skew-symmetric operator. Finally, the disturbance vector is modeled as a finite sum of sinusoidal modes,
\begin{equation}
    \delta\theta_{q,d}(t)
    =
    \sum_{k=1}^{N_q}
    a_{q,k}
    \sin\!\left(
        2\pi f_{q,k} t + \phi_{q,k}
    \right),
    \qquad q\in\{x,y,z\}.
\end{equation}
Here, $f_{q,k}$ are the disturbance frequencies for axis $q$, $a_{q,k}$ are the corresponding amplitudes, and $\phi_{q,k}$ are constant phase offsets. In the simulations presented here, $N_z=2$ modes are used to generate a simplified representative low-frequency disturbance environment without reproducing the full measured disturbance spectrum. The phase offsets $\phi_k$ can be sampled uniformly from $[0,2\pi)$, and amplitudes are fixed for each simulation scenario using a calibration derived from flight observations.\\

\subsubsection{Disturbance Calibration}

SuperBIT flight data suggest stronger coupling between RSC motion and passive FSC motion than would be predicted by a rigid-body geometric model. To quantify this discrepancy, we first derive the expected relationship between RSC and passive FSC image motion.

Under a rigid-body assumption, with the FSM locked onto one guide star and for small angular displacements, a simple estimate of displacement at a passive FSC is
\begin{equation}
    \Delta x_{fsc}\ \text{[mm]} = \ell'\ \text{[mm]}\cdot \delta\theta_{rsc}\ \text{[rad]},
\end{equation}
where $\ell'$ is the separation between the active and passive guide stars on the focal plane. Expressed in pixel coordinates, the expression rearranges to
\begin{equation}
    \Delta x_{fsc}\ \text{[px]}
    =
    \alpha
    \frac{\ell'}{f_{RSC}}
    \Delta x_{rsc}\ \text{[px]}
    \approx
    \underbrace{0.135}_{k_{\mathrm{geo}}}\alpha
    \Delta x_{rsc}\ \text{[px]},
\end{equation}
where $k_{\mathrm{geo}}$ is the geometric coupling coefficient, $f_{RSC}$ is the focal length of the RSC, and $\alpha=\zeta_{RSC}/\zeta_{FSC}$ is the pixel-size ratio. For FSC1, $\alpha_1=1$, while for FSC2, $\alpha_2\approx0.76$. In this estimate, $\ell'$ is approximated by the separation between the FSC centers, $\ell$. With no FSM control, the amplitude of $\Delta x_{FSC}$ is divided by $2$, but so is the lever arm $\ell'$, preserving the ratio.

\begin{figure}[!h]
    \centering
    \includegraphics[width=0.6\linewidth]{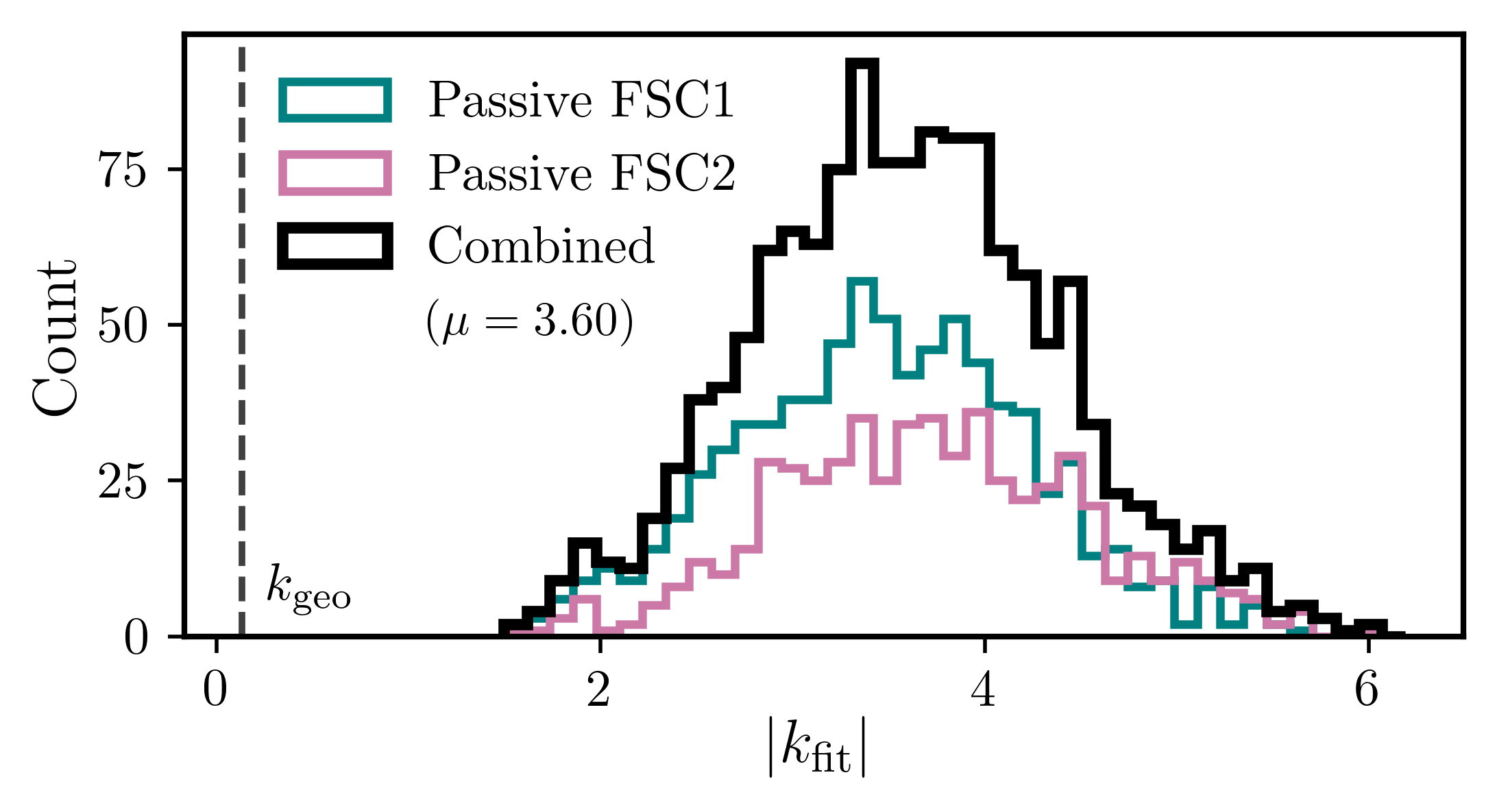}
    \caption{
    Distribution of the measured coupling magnitude $|k_{\mathrm{fit}}|$ from 2023 SuperBIT flight data between the passive FSC and RSC after correcting FSC2 by the pixel-size factor $\alpha$ ($n=1370$). The dashed line shows the rigid-body geometric prediction $k_{\mathrm{geo}}$. The combined distribution has $\mu=3.60$ and $\sigma=0.83$.
    }
    \label{fig:khistogram}
\end{figure}

However, flight data consistently exhibit larger passive FSC motion than predicted by this geometric relationship. The measured distribution of passive FSC-to-RSC coupling coefficients is shown in Figure \ref{fig:khistogram}. For all exposures with two star cameras, as flagged by Eq. \ref{eq:flag1}, and satisfying
\begin{equation}
\Delta x_{\mathrm{RSC}} > 2.5~\mathrm{px},
\qquad
|\rho_{x,\mathrm{RSC},x,\mathrm{FSC}}| > 0.6,
\label{eq:flag2}
\end{equation}
which selects exposures with high roll motion and roll leakage, the best-fit coupling coefficient $k_{\mathrm{fit}}$ is computed as
\begin{equation}
    k_{\mathrm{fit}}
    =
    \arg\min_{k \in \mathbb{R}}
    \left\|
        \tilde{x}_{fsc}-k\,x_{rsc}
    \right\|_2^2,
\end{equation}
where $\tilde{x}_{fsc}=x_{fsc}/\alpha_i$. Of the $2348$ exposures selected by Eq. \ref{eq:flag1}, $1370$ ($\sim58.3\%$) satisfy the high-roll criteria of Eq. \ref{eq:flag2}.

The resulting distribution of $|k_{\mathrm{fit}}|$ exceeds the rigid-body prediction $k_{\mathrm{geo}}$, by about a factor of $\sim 27$. This indicates that passive focal-plane motion is amplified by effects not captured by the simple by field rotation assumptions. A likely contribution is field-dependent tip-tilt projection, through which roll-driven FSM commands can amplify the focal-plane motion associated with field rotation. Other possible contributions include torsional flexure and other unmodeled roll dynamics. Nevertheless, the correlation between the RSC and passive FSC remains high despite this amplification, suggesting that the observed motion is still predominantly roll-driven. The amplification is consistently observed across the flight data and is therefore treated empirically in the present analysis. To account for this excess response, we use the mean $|k_{\mathrm{fit}}|$ of the combined distribution to scale the simulated model in the numerical results that follow.\\

\section{Numerical Results}
\vspace{3pt}
\label{sec:Numerical_results}

Numerical simulations are performed first using the SuperBIT design, with dimensions provided in Table \ref{tab:optical_geometry}. Three cases are considered. First, an open-loop roll-only disturbance is used to illustrate the focal-plane response of the optical model. Second, single-guide-star and multi-guide-star guidance are compared using the SuperBIT optical geometry. Finally, the same comparison is repeated using the GigaBIT optical geometry. To isolate the effect of roll leakage on science-camera image, pitch and cross-pitch disturbances are neglected in these simulations by setting $\delta\theta_{x,d}=\delta\theta_{y,d}=0$. Implications of this will be discussed later.

\begin{table}[!h]
\caption{Optical geometry used for the SuperBIT and GigaBIT simulations.}
\vspace{3pt}
\label{tab:optical_geometry}
\centering
\renewcommand{\arraystretch}{1.05}
\begin{tabular}{lcc}
\hline
\rule{0pt}{2.2ex}
\textbf{Parameter} &
\textbf{SuperBIT} &
\textbf{GigaBIT\tablefootnote{Approximate values representative of the scale of the GigaBIT optical geometry \cite{AMiles2026}.}} \\
\hline
FSC separation $\ell$ [mm] & 40.6 & 150 \\
FSM--focal plane distance $d$ [mm] & 180 & 610 \\
\hline
\end{tabular}
\end{table}

\begin{figure}[h!]
\centering
\includegraphics[width=\linewidth]{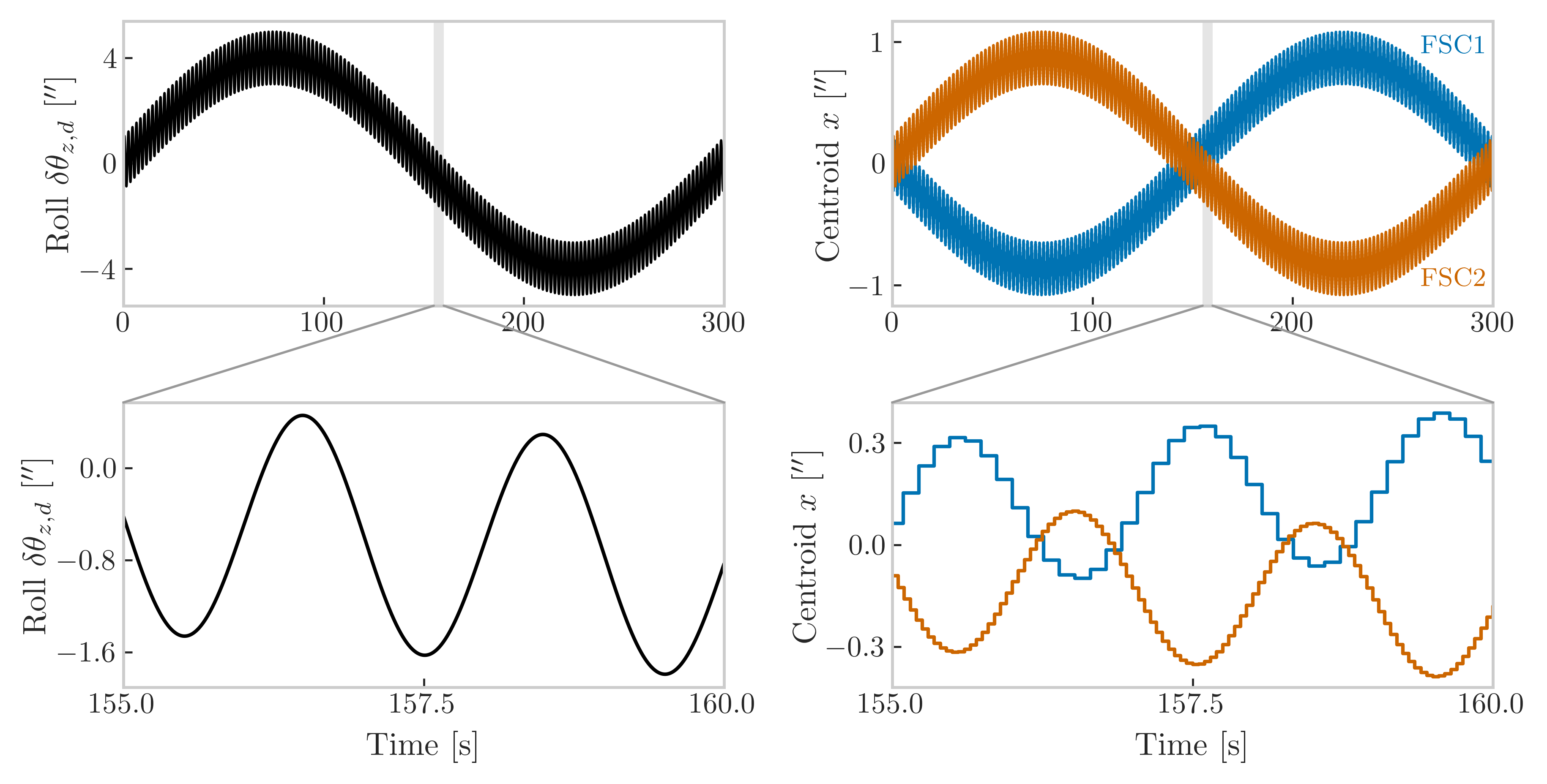}
\caption{
Simulation response to a roll-only disturbance with the FSM inactive. The left column shows the injected roll disturbance, while the right column shows the resulting focal-plane motion measured by the two FSCs. The lower panels show a magnified view of the shaded interval. The equal-and-opposite centroid motion is a consequence of the symmetric placement of the FSCs about the boresight.
}
\label{fig:sim_result_1}
\end{figure}

We begin by simulating a roll-only disturbance with FSM inactive to illustrate the behavior of the optical model. We inject a representative low-frequency roll disturbance consisting of two sinusoidal modes at $0.5\,\mathrm{Hz}$ and $1/300\,\mathrm{Hz}$ with amplitudes of $1''$ and $4''$, respectively, with $\boldsymbol{\Sigma}_v=\mathbf{0}$ and synchronous ($\sigma_{f,1}=\sigma_{f,2}=0\,\mathrm{Hz}$) FSC update rates of $f_1=8\,\mathrm{Hz}$ and $f_2=20\,\mathrm{Hz}$. The result is shown in Figure \ref{fig:sim_result_1}, where the roll disturbance is transmitted directly to the focal plane, producing equal-and-opposite centroid motion on the two FSCs, as we would expect. We plot the $x$-axis centroid coordinate, since the corresponding $y$-axis response is negligible for a pure roll disturbance.

\subsection{Estimator Comparison}

The objective of the FSM is to improve image stability over the science field. For pitch and cross-pitch disturbances, this is accomplished by reducing common translational image motion across the focal plane. In the roll-only case considered here, however, the disturbance cannot be removed. Instead, the FSM can only change how the roll-induced image motion is distributed across the focal plane by shifting the center of rotation. The best achievable outcome is therefore to place the center of rotation near the science field and minimize science-camera image smear.

We next evaluate the single-guide-star and multi-guide-star estimators defined in Section \ref{sec:Measurement_Estimation} using the SuperBIT layout. For the single-guide-star case, FSC2, the higher-rate camera, is selected as the active guide star. All simulation parameters are shown in Table \ref{tab:sim_parameters}. Here, we introduce some amount of sensor noise, as well as make the star cameras asynchronous, as explained in Section \ref{sec:Measurement_Estimation}. The injected roll jitter amplitudes, which are the same as the left quadrant of Figure \ref{fig:sim_result_1}, are intentionally larger than expected flight levels to better highlight differences between the estimation architectures. 

\begin{table}[h]
\caption{Simulation parameters used, unless otherwise stated.}
\vspace{3pt}
\label{tab:sim_parameters}
\centering
\renewcommand{\arraystretch}{1.05}
\begin{tabular}{llc}
\hline
\rule{0pt}{2.2ex}
\textbf{Parameter} &
\textbf{Variable} &
\textbf{Value} \\
\hline
FSC1 mean update rate & $\bar{f}_1$ & $8\,\mathrm{Hz}$ \\
FSC1 update-rate std. & $\sigma_{f,1}$ & $2\,\mathrm{Hz}$ \\
FSC2 mean update rate & $\bar{f}_2$ & $20\,\mathrm{Hz}$ \\
FSC2 update-rate std. & $\sigma_{f,2}$ & $5\,\mathrm{Hz}$ \\
FSC1 centroid noise & $\sigma_{v,1}$ & $0.04\,\mathrm{px}$ \\
FSC2 centroid noise & $\sigma_{v,2}$ & $0.07\,\mathrm{px}$ \\
Multi-star nominal weights & $\tilde{w}_1,\tilde{w}_2$ & $1,\ 1$ \\
Multi-star decay time & $\tau_d$ & $1.0\,\mathrm{s}$ \\
FSM time constant & $\tau$ & $0.05\,\mathrm{s}$ \\
Proportional gain, $x$ axis & $\kappa_{p,x}$ & $1.00\times10^{-4}\,\mathrm{rad\,mm^{-1}}$ \\
Proportional gain, $y$ axis & $\kappa_{p,y}$ & $6.94\times10^{-5}\,\mathrm{rad\,mm^{-1}}$ \\
Integral gain, $x$ axis & $\kappa_{i,x}$ & $1.00\times10^{-4}\,\mathrm{rad\,mm^{-1}\,s^{-1}}$ \\
Integral gain, $y$ axis & $\kappa_{i,y}$ & $1.00\times10^{-4}\,\mathrm{rad\,mm^{-1}\,s^{-1}}$ \\
Empirical FSC gain &  & $58.39$ \\
Roll disturbance amplitudes &
$a_{z,1},\ a_{z,2}$ &
$1'',\ 4''$ \\
Roll disturbance frequencies & $f_z$ & $0.5\,\mathrm{Hz},\,1/300\,\mathrm{Hz}$ \\
Roll disturbance phases & $\phi_z$ & $0,\ 0$ \\
Pitch \& x-pitch disturbance & $\delta\theta_{x,d}$, $\delta\theta_{y,d}$ & $0'',\ 0''$\\
Simulation duration & $T$ & $300\,\mathrm{s}$ \\
Simulation time step & $\Delta t$ & $0.01\,\mathrm{s}$ \\
\hline
\end{tabular}
\end{table}

\begin{figure}[h]
\centering
\includegraphics[width=\linewidth]{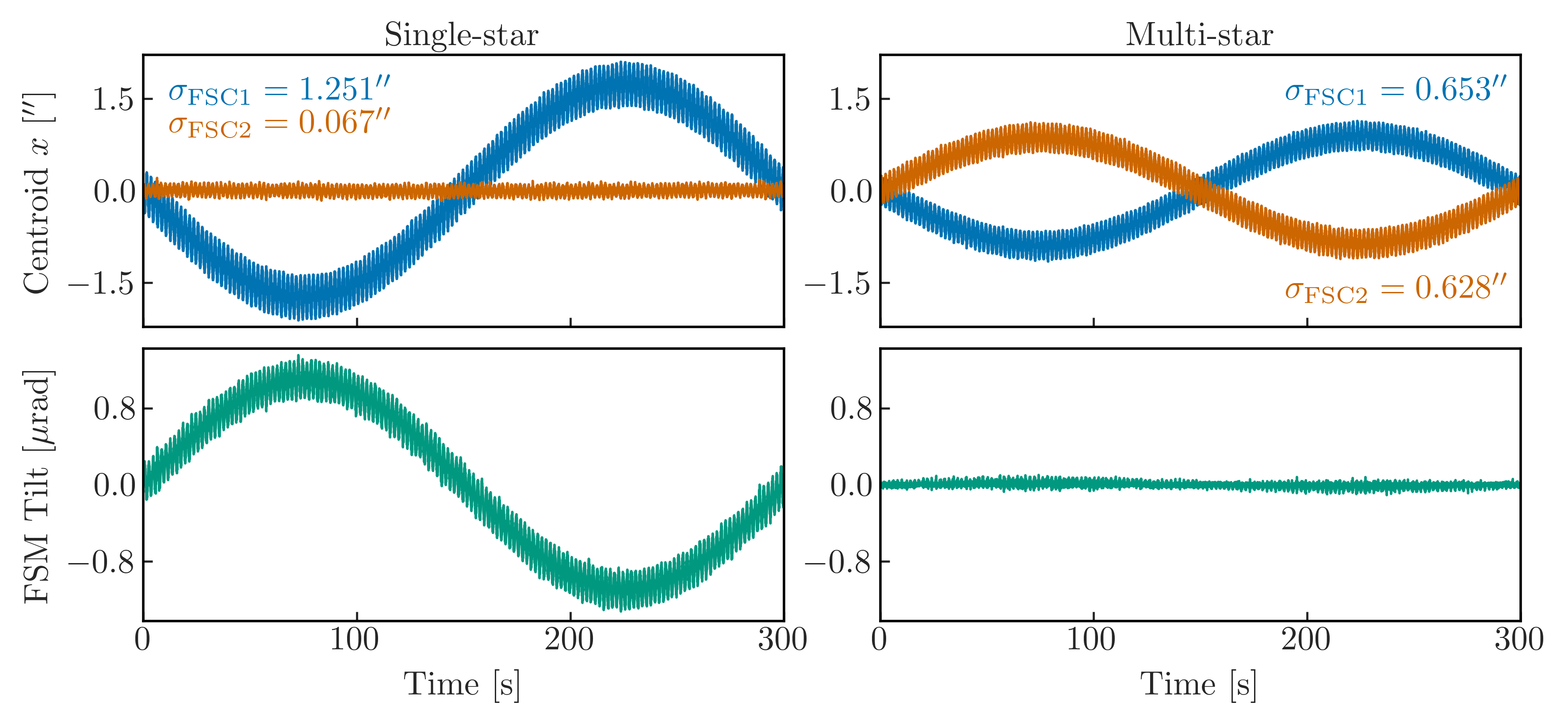}
\caption{
Closed-loop response of single-guide-star and multi-guide-star estimation under a roll-only disturbance. The top row shows FSC centroid motion and the bottom row shows FSM tilt commands.
}
\label{fig:single_star_control}
\end{figure}

Figure \ref{fig:single_star_control} compares the resulting closed-loop responses. The single-guide-star controller successfully stabilizes the active guide star, reducing the FSC2 centroid motion to a residual standard deviation of $0.067''$, while the passive FSC1 experiences substantially larger image motion than in the open-loop case of Figure \ref{fig:sim_result_1}. This reproduces the roll-leakage behavior discussed previously.

In contrast, the multi-guide-star estimator produces nearly equal residual image motion at both FSCs, with standard deviations of $0.653''$ and $0.628''$ for FSC1 and FSC2, respectively. Rather than optimizing image stability at a single field location, the estimator distributes the residual roll motion more uniformly across the focal plane, namely by moving the center of rotation back to the center of the science camera. As a result, the FSM commands become nearly zero, since the boresight lies midway between the two FSCs and coincides with the center of the science field. Small nonzero FSM commands arise from the asynchronous camera updates and the age-weighted measurement fusion used by the esxtimator. More general focal-plane layouts, for example non-symmetric configurations such as GigaBIT, can be accommodated through the adjustable nominal camera weights of Eq.~\ref{eq:weights}, allowing the center of correction to be biased toward the science field.

For scenarios in which one guide camera updates significantly faster than the other, the centroid estimate becomes increasingly weighted toward the more recent measurements from that camera. In the limit of a large update-rate difference, the estimator approaches single-guide-star behavior and the center of rotation shifts toward the faster guide star. The multi-guide-star solution therefore sacrifices perfect stabilization at the guide-star location in exchange for improved image stability over the science field as a whole.

\begin{figure}[h]
\centering
\includegraphics[width=0.82\linewidth]{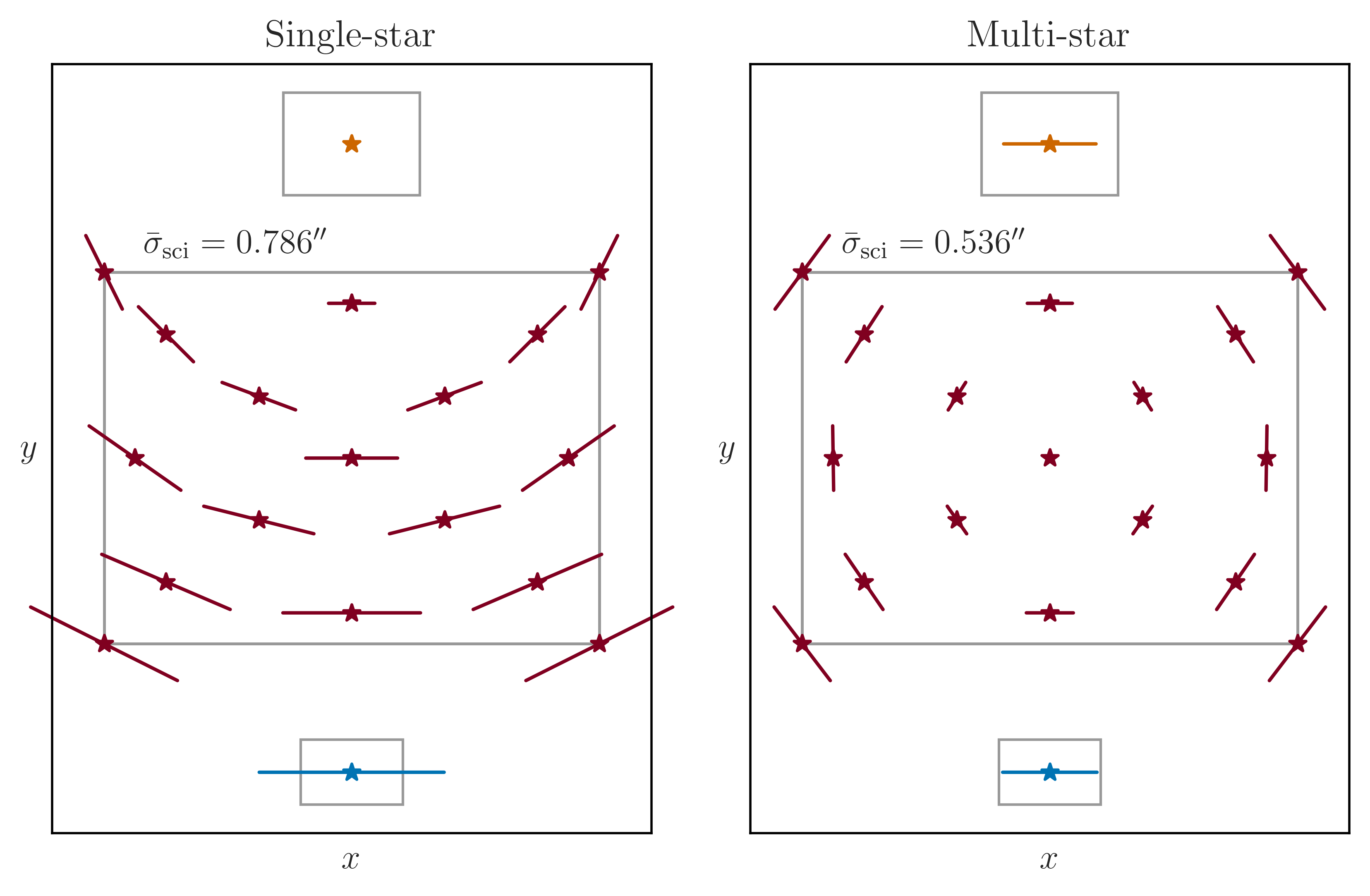}
\caption{
Image trails across the science field computed from the simulated focal-plane trajectories for the single-guide-star and multi-guide-star estimators. Line lengths are scaled by a factor of 200 for visibility. Using both guide stars reduces the average science-field image motion from $0.786''$ to $0.536''$.
}
\label{fig:science_smear}
\end{figure}

To quantify science performance, we use the average RMS image motion across the science camera focal plane. Simulated stars are placed at 16 representative locations spanning the science field of view, including the field corners, and the resulting RMS image motions are averaged to produce a single image-quality metric,
\begin{equation}
    \bar{\sigma}_{\mathrm{sci}}
    =
    \frac{1}{16}
    \sum_{s=1}^{16}
    \sqrt{
        \sigma_{x,s}^2
        +
        \sigma_{y,s}^2
    },
\end{equation}
where $\sigma_{x,s}$ and $\sigma_{y,s}$ are the RMS image motions of science-field star $s$ along the focal-plane axes. Figure \ref{fig:science_smear} shows the resulting image motion across the science field. The single-guide-star estimator produces elongated image trails across much of the focal plane, resembling the roll-leakage patterns observed in the SuperBIT flight data (Figure \ref{fig:figure2}). Using both guide stars shortens these trails and reduces the average science-field image motion from $0.786''$ to $0.536''$. \\

\subsection{Application to GigaBIT}
The previous results were obtained using the SuperBIT focal-plane geometry. GigaBIT employs a substantially larger focal plane, increasing the separation between guide stars and the science field. Since roll-induced image motion scales with distance from the center of rotation, the impact of guide-star geometry is expected to be more pronounced.

The same disturbance realization and estimator configurations are then simulated using the representative GigaBIT geometry from Table \ref{tab:optical_geometry}. Although the planned GigaBIT focal plane is more complex and annular \cite{AMiles2026}, we model it here using a scaled version of the SuperBIT geometry with representative dimensions. The control gains are scaled by the ratio $180/610$ to account for the larger FSM-to-focal-plane distance in the GigaBIT optical design. Finally, the GigaBIT detector sizes are used, corresponding to an IMX811 science camera and IMX540 tracking cameras, and fake stars are placed in the same relative positions as with the SuperBIT design.

Results for both geometries are summarized in Table \ref{tab:science_rms_comparison}. For the SuperBIT geometry, the multi-guide-star estimator reduces the average science-field image motion by $31.8\%$ relative to the single-guide-star architecture. While this improvement is noticeable, the single-guide-star solution still provides reasonable image stability across much of the science field. The improvement is much larger for the GigaBIT geometry, where the same estimator reduces the image motion by $77.4\%$. In both cases, the multi-guide-star solution produces lower science-field image motion despite not minimizing the motion at any individual guide star.

\begin{table}[!h]
\caption{Science-field average RMS image motion for SuperBIT and GigaBIT.}
\vspace{3pt}
\label{tab:science_rms_comparison}
\centering
\renewcommand{\arraystretch}{1.05}
\begin{tabular}{lccc}
\hline
\rule{0pt}{2.2ex}
\textbf{Geometry} &
\textbf{Single-star $\bar{\sigma}_{\mathrm{sci}}$ [$''$]} &
\textbf{Multi-star $\bar{\sigma}_{\mathrm{sci}}$ [$''$]} &
\textbf{Reduction} \\
\hline
SuperBIT & $0.786$ & $0.536$ & {\color{seagreen}\textbf{31.8\%}} \\
GigaBIT  & $2.343$ & $0.529$ & {\color{seagreen}\textbf{77.4\%}} \\
\hline
\end{tabular}
\end{table}

The larger improvement observed for GigaBIT reflects the increased sensitivity of wide-field systems to roll-induced image motion. As focal-plane area increases, the limitations of single-guide-star control become more apparent. For SuperBIT, single-guide-star fine guidance was sufficient for science observations. For GigaBIT, however, the image-quality penalty associated with single-guide-star control becomes much larger, making the multi-guide-star solution considerably more effective. We note that this comparison does not account for the higher effective guide-star centroid update rate expected for GigaBIT, which may further improve fine-guidance performance. Future work should apply the same framework to the planned GigaBIT focal-plane layout, which is more complex than the simplified geometry used here. \\

\section{Conclusion}
\vspace{1pt}

This work has characterized roll leakage as a systematic limitation of single-guide-star fine-guidance architectures in balloon-borne telescopes, and evaluated multi-star estimation as a mitigation strategy. Three main contributions were presented. First, analysis of SuperBIT 2023 science flight data showed that residual focal-plane image motion is frequently correlated with boresight roll across a large population of science exposures. The passive FSC centroid motion tracked RSC roll measurements in the majority of exposures that satisfied quality criteria, confirming that roll leakage is an inherent feature of the single-guide-star fine-guidance architecture.  Second, a closed-loop simulation framework combining optical ray tracing, asynchronous FSC measurements, and FSM control was developed to study the coupling between boresight roll, guide-star geometry, and science-field image quality. The framework reproduces the roll-leakage behavior observed in flight data and provides a controlled environment for comparing guidance architectures under representative disturbance conditions. Third, single-star and multi-star estimation architectures were compared using representative SuperBIT and GigaBIT optical geometries. For SuperBIT, multi-star estimation reduced average science-field image motion by $31.8\%$ relative to single-star guidance. For GigaBIT, the same approach yielded a $77.4\%$ reduction, reflecting the greater sensitivity of wide-field systems to roll-sourced focal-plane motion. The benefit of multi-star fine guidance therefore scales with focal-plane size, making it increasingly relevant as balloon-borne observatories grow in aperture and field of view.

These results motivate the adoption of multi-star fine-guidance architectures for GigaBIT, where roll leakage poses a substantially greater challenge than for SuperBIT. Future work should incorporate gyroscope measurements to improve estimation between camera updates, replace the simplified focal-plane geometry used here with the planned GigaBIT layout, and investigate the effects of pitch and cross-pitch disturbances on science-field image motion. Multi-star estimation should also be evaluated as part of upcoming GigaBIT flight-testing efforts. Improving coarse roll rejection at the gimbal level remains a complementary approach and, combined with multi-star fine guidance, provides a path toward meeting GigaBIT's 0.020 arcsec image-stability requirement.\\

\acknowledgments 
The SuperBIT 2023 science flight could only have been possible with the contribution of many people. The SuperBIT pre-flight integration campaign took place at the NASA Columbia Scientific Balloon Facility (CSBF) under contract from NASA’s Balloon Program Office (BPO). The launch from the W\=anaka airport in New Zealand was provided by NASA, with support by NASA CSBF personnel. The US team acknowledges support from the NASA APRA grant 80NSSC22K0365. The research was carried out in part through the Jet Propulsion Laboratory, California Institute of Technology, under a contract with the National Aeronautics and Space Administration (80NM0018D0004). Canadian team members acknowledge support from the Canadian Institute for Advanced Research (CIFAR), Natural Science and Engineering Research Council (NSERC), and the Canadian Space Agency (CSA). UK coauthors acknowledge funding from Durham University’s Astronomy Projects Award, STFC (grants ST/P000541/1, ST/V005766/1 and ST/X001075/1), and UKRI (grant MR/X006069/1). These works were supported by a grant (10.69777/2005729) from the Fonds de recherche du Québec. GigaBIT is additionally supported by the Canadian Foundation for Innovation (CFI) and the Ontario Research Fund (ORF). AI tools were used for editorial assistance in this work.

\newpage
\bibliography{report} 
\bibliographystyle{spiebib} 

\end{document}